\documentclass[11pt]{article}
\usepackage[pctex32]{graphics}
\begin{document}
\vspace{1cm}
\begin{center}
~\\
{\bf  \Large  KK6 from M2 in BLG}
\vspace{2cm}

                      Wung-Hong Huang\\
                       Department of Physics\\
                       National Cheng Kung University\\
                       Tainan, Taiwan\\

\end{center}
\vspace{2cm}
\begin{center}{\bf  \Large ABSTRACT } \end{center}
We study the possibility that the Kaluza-Klein monopole (KK6) world-volume action may be obtained from the multiple membranes (M2) action which is described by  BLG theory.   We  first point out that the infinite dimensional Lie 3-algebra based on the Nambu-Poisson structure could not only provide three dimensional manifolds to allow M5 from M2, which was studied by previous authors, but also provide five dimensional manifolds to allow KK6 from M2.  We next present a possible way that the U(1) field on KK6 world-volume action could be produced form the gauge potential in BLG theory.
\vspace{4cm}
\begin{flushleft}
*E-mail:  whhwung@mail.ncku.edu.tw\\
\end{flushleft}
\newpage
\section{Introduction}
The work of Bagger, Lambert and equivalently Gustavsson (BLG
theory) [1-4] had found a theory for multiple M2 branes using a wonderful
algebraic structure, Lie 3-algebra [5-7].   The bosonic part Lagrangian for multiple M2-branes in BLG theory contains kinetic term, potential term and Chern-Simons term, which are 
$$L = -{1\over 2}D^\mu X^{aI}D_\mu X_a^{I}-{1\over12 }Tr[X^{I},X^{J},X^{K}][X^I,X^J,X^K]$$
$$+{1\over2}\epsilon^{\mu\nu\lambda}\left(f^{abcd}A_{\mu ab}\partial_\nu A_{\lambda cd}+ {2\over3}f^{cda}_{~~~~g} f^{efgb}A_{\mu ab}A_{\nu cd}A_{\lambda ef}\right).\eqno{(1.1)}$$
The covariant derivative $D_\mu X_a^I$ is defined by $D_\mu X_a^I = \partial_\mu X_a^I - A_{\mu~a}^{~b}X_b^I$ in which $A_{\mu~a}^{~b}$ is a gauge field with two algebraic indices.  The indices I, J, K run in $1, \cdot \cdot\cdot, 8$, which specify the transverse directions of M2-brane; $\mu$, $\nu$ and $\lambda$ run in 0, 1, 2, which describe the longitudinal directions. The indices $a, b,\cdot \cdot\cdot,f$ take values in $1, \cdot \cdot\cdot, {\cal D}$ where ${\cal D}$ is the number of generators $T^a$ of the Lie 3-algebra specified by a set of structure constants $f^{abc}_{~~~d}$ in a trilinear antisymmetric product
$$[T^a, T^b, T^c] = f^{abc}_{~~~d}~T^d.\eqno{(1.2)}$$
It supposes that  there is trace-form that provides a metric
$$h^{ab} = Tr(T^a, T^b),\eqno{(1.3)}$$
which allows us to raise and lower indices: $f^{abcd} = f^{abc}_{~~~e}h^{ed}$.

The consistency condition of Lie 3-algebra is that it must satisfy the
so-called fundamental identity [1] : 
$$[T^a, T^b, [T^c, T^d, T^e]] = [[T^a, T^b, T^c], T^d, T^e]
+ [T^c, [T^a, T^b, T^d], T^e] + [T^c, T^d, [T^a, T^b, T^e]].\eqno{(1.4)}$$
and 
$$Tr([T^a, T^b, T^c], T^d) + Tr(T^c, [T^a, T^b, T^d]) = 0.\eqno{(1.5)}$$
The Lie 3-algebra used in the original BLG model is four-dimensional vector space, denoted as $A_4$, which is equivalent to one based on $SU(2) \times SU(2)$ and describes only two M2-branes [8].

It is well-known that Nambu algebras are a particular, infinite-dimensional case of n-Lie algebras.  Their n-bracket is provided by the Jacobian determinant of n functions [5].  The use of the Nambu bracket in the context of the BLG model was initially mentioned in [3] and studied extensively in [9,10]. The novelty introduced by the Nambu bracket is that the infinite-dimensional Lie algebra turns out to be the volume preserving diffeormorphisms group. 

  More precisely, the scalar field $X^I$ and gauge field $A_{\mu a}$ in BLG are expanded in terms of a basis $\chi^a(y)$ of  Nambu-Poisson bracket [10]
$$X^I(x,y) = \sum_{a}X_a^I(x)\chi^a(y).\eqno{(1.6)}$$
$$A_{\mu a}(x,y) = \sum_{a}A_{\mu ab}(x)\chi^b(y).\eqno{(1.7)}$$
The coordinate $x$ is the 3 dimensional  world volume of M2 brane while coordinate $y$ is the 3 dimensional internal spaces coming from 3-algebra. In this approach Ho and Matsuo [9] found M5 world-volume action from M2 action in BLG theory.  This means that they consider an 3 dimensional internal spaces in the world volume of M2 brane and find a six dimensional theory which has some desired properties of an M5 brane.  In particularly,  they had found the action of a self-dual two form gauge field living on the world volume of M5 brane. 

   Note that the decomposability [6] of the Nambu-Poisson bracket tells us that, locally one can always choose 3 coordinates $(\mu,\nu,\lambda)=(x_1, x_2, x_3)$ in terms of which the bracket is simply 
$$[f(x_1,x_2,x_3,\cdot\cdot\cdot),g(x_1,x_2,x_3,\cdot\cdot\cdot),h(x_1,x_2,x_3,\cdot\cdot\cdot)]= \epsilon^{\mu\nu\lambda}\partial_\mu f \partial_\nu g \partial_\lambda h.\eqno{(1.8)}$$
At first sight, the rest of the coordinates ($x_i$, for $i > 3$) will not induce derivative components and there can never be more than 3 of the $x_i$  to turn into covariant derivatives in studying M5 from M2.  Thus the decomposability of the Nambu-Poisson bracket is the mathematical basis of why there are no other Mp-branes with $p\ne 5$, as mentioned in [10].
\\

However, in M-theory there are Kaluza-Klein monopole (KK6) object which is a six dimensional object.  As discussed in [11], a special feature of the Kaluza-Klein monopole is that one of its four transverse directions corresponds to the isometry direction in the Taub-NUT space of  the monopole [12]. Since the monopole cannot move in this direction one should not associate a physical worldvolume scalar to it.  It was argued in [13] that  M2-brane can only intersect with  KK6  over a 0-brane such that one of the worldvolume directions of the M-2-brane coincides with the isometry direction z of the Taub-NUT space
\\
$$ \left(0|M2,KK6\right)=\{\begin{tabular}{l|rcrrrrrrrr}
$\times$    & $\times$ & $\times$&-&-&-&-&-&-&-&-  \\
$\times$  & - & $z$&$\times$&$\times$&$\times$&$\times$&$\times$&$\times$&- &-
\end{tabular}$$
\\
In the previous study [10], as M2-brane can intersect with  M5  over a 1-brane, i.e. $(1| M2,M5)$, we need extra 3 dimensional internal spaces from 3-algebra in M2 brane to have 5 space dimensions in the world volume of M5.  The extra 3 dimensional internal spaces is that in Nambu-Poisson bracket.    However, in considering $(0|M2,KK6)$ we need extra 5 dimensional internal spaces from 3-algebra in M2 brane to have 6 space dimensions in the world volume of KK6.  This seems to conflict to the relation (1.8). 
\\

   In section 2 we  point out that the infinite dimensional Lie 3-algebra based on the Nambu-Poisson structure could provide five dimensional manifolds to allow KK6 from M2. In section 3  we present a possible way that the U(1) field on KK6 world-volume action [14] could be produced form the gauge potential in BLG action.
\section {Brane Worldvolume from  N-Lie Algebras}
{\bf Simple observation}. Let us first consider the simplest case : 2-Lie algebra.  In this case the fundamental identity (1.4) is the Jacobi identity.
$$[A, [B, C]] = [[A, B], C]+ [B, [A, C]].\eqno{(2.1)}$$
It is a simple work to prove that the following three representations
$$ [A,B]_{(xy)}= \partial_x A(x,y,z) \partial_y B(x,y,z)-\partial_y A(x,y,z) \partial_x B(x,y,z),\eqno{(2.2)}$$
$$ [A,B]_{(yz)}= \partial_y A(x,y,z) \partial_z B(x,y,z)-\partial_z A(x,y,z) \partial_y B(x,y,z),\eqno{(2.3)}$$
$$ [A,B]_{(zx)}= \partial_z A(x,y,z) \partial_x B(x,y,z)-\partial_x A(x,y,z) \partial_z B(x,y,z),\eqno{(2.4)}$$
all automatically satisfies the Jacobi identity. 

  In fact, we can furthermore prove that the representation
$$ [A(x,y,z),B(x,y,z)]= \sum_{\mu=x,y,z \atop\nu=x,y,z}\epsilon^{\mu\nu}\partial_\mu A(x,y,z) \partial_\nu B(x,y,z)= [A,B]_{(xy)}+[A,B]_{(yz)}+[A,B]_{(zx)},\eqno{(2.5)}$$
also satisfies the Jacobi identity.  In this case of 2-algebra we see that the property of  decomposability does not constrain to appearing derivatives no more than 2 coordinates, contrasts to the previous belief. 

Therefore it is naturally to suspect that, in the case of 3-algebra with $(\mu,\nu,\lambda)=(x_1, x_2, x_3,x_4)$  the representation 
$$[f(x_1,x_2,x_3,x_4),g(x_1,x_2,x_3,x_4),h(x_1,x_2,x_3,x_4)]=\epsilon^{\mu\nu\lambda}\partial_\mu f \partial_\nu g \partial_\lambda h,\eqno{(2.6)}$$
will satisfies the fundamental identity.   In fact, we can easily see this property from the following theorem.

\subsection{Theorem 1}
 {\bf Theorem 1 : (n-k) Lie algebras from n-Lie algebras}. Let $G$ be an arbitrary n-Lie algebra and $k$ fixing elements $A_1,\cdot\cdot\cdot,A_k\epsilon ~G$ in its n-bracket.  Define  the  (n-k)-linear and fully antisymmetric (n-k) bracket by
$$[X_1,X_2, . . . ,X_{n-k}]_{n-k} \equiv  [A_1,\cdot\cdot\cdot,A_k,X_1,X_2, . . . ,X_{n-k}]_n.\eqno{(2.7)}$$
Then, the (n-k)-bracket defined above satisfies the fundamental identity (1.4).
\\
\\
{\it Proof :} Clearly, with $A_i$ fixed, the n-bracket implies the equality
$$[A_1,\cdot\cdot\cdot,A_k,X_1,X_2, . . . ,X_{n-k-1}, [A_1,\cdot\cdot\cdot,A_k, Y_1, Y_2, . . . , Y_{n-k}]]_n $$
$$= [[A_1,\cdot\cdot\cdot,A_k,X_1,X_2, . . . ,X_{n-k-1},A_1],A_2,\cdot\cdot\cdot,A_k, Y_1, Y_2, . . . , Y_{n-k}]]_n$$
$$ + [A_1,[A_1,\cdot\cdot\cdot,A_k,X_1,X_2, . . . ,X_{n-k-1},A_2],A_3,\cdot\cdot\cdot,A_k, Y_2, . . . , Y_{n-k}]]_n+\cdot\cdot\cdot$$
$$+[A_1,\cdot\cdot\cdot,A_k,[A_1,\cdot\cdot\cdot,A_k,X_1,X_2, . . . ,X_{n-k-1},Y_1], Y_2, . . . , Y_{n-k}]]_n $$
$$ +\cdot\cdot\cdot+  [A_1,\cdot\cdot\cdot,A_k,Y_1, . . . , Y_{n-k-1},[A_1,\cdot\cdot\cdot,A_k,X_1,X_2, . . . ,X_{n-k-1},Y_{n-k}]]_n.\eqno{(2.8)}$$
Using definition (2.7) above relation becomes
$$[X_1,X_2, . . . ,X_{n-k-1}, [Y_1, Y_2, . . . , Y_{n-k}]]_{n-k} = 0+0+\cdot\cdot\cdot
 + [[X_1,X_2, . . . ,X_{n-k-1},Y_1], Y_2, . . . , Y_{n-k}]]_{n-k}$$
$$+\cdot\cdot\cdot+  [Y_1, . . . , Y_{n-k-1},[X_1,X_2, . . . ,X_{n-k-1},Y_{n-k}]]_{n-k},\eqno{(2.9)}$$
which just is the equality of (n-k)-bracket.   This extends the theorem of $k=1$ in [7].
\\
\\
{\it Application 1:} For the case of k=1: Now, using the decomposability [6] of the Nambu-Poisson bracket of 3-algebra we can find the 2-algebra from above theorem by define
$$[A, B] \equiv [x+y+z, A, B] = \sum_{(\mu\nu)=(x,y),(y,x),\atop (yz),(zy),(zx)(xz),}\epsilon^{\mu\nu}\partial_\mu A(x,y,z) \partial_\nu B(x,y,z).\eqno{(2.10)}$$
We then obtain the representation  (2.5).  Of course we can see that the representations
$$ [x\pm y\pm z, A, B] = [A,B]_{(xy)}\pm [A,B]_{(yz)}\pm [A,B]_{(zx)},\eqno{(2.11)}$$
also satisfy the fundamental identity.  The difference between them is the order of (x,y,z) and may be unable to affect the physical result.
\\
\\
{\it Application 2:} For the case of k=1: Now, using the decomposability [6] of the Nambu-Poisson bracket of 4-algebra we can find the 3-algebra from above theorem by fixing $A_1=x_1+x_2+x_3+x_4$.  Then
$$[f, g, h] \equiv [x_1+x_2+x_3+x_4, f, g, h] =  \sum_{(\mu\nu\lambda)}\epsilon^{\mu\nu\lambda}\partial_\mu f \partial_\nu g \partial_\lambda,\eqno{(2.12)}$$
and we obtain the representation  (2.6).  Thus, we have extra  4 internal spaces from 3-algebra in BLG theory.  
\\
\\
{\it Application 3:} For the case of k=2  we can use the decomposability [6] of the Nambu-Poisson bracket of 5-algebra to find the 3-algebra from above theorem by fixing $A_1=x_1+x_2+x_3$, $A_2=x_4+x_5$.  Then
$$[f, g, h] \equiv [x_1+x_2+x_3, x_4+x_5, f, g, h] = \sum_{(\mu\nu\lambda)}^{(x_4+x_5)\atop (x_1,x_2,x_3)} \epsilon^{\mu\nu\lambda}\partial_\mu f \partial_\nu g \partial_\lambda h.\eqno{(2.13)}$$
Thus, we have extra 5 dimensional internal spaces from 3-algebra in BLG theory, which may be used to describe KK6.   Note that, above choice of $A_1$ and $A_2$ seems that it has only symmetry $O(3)\times O(2)$ while not O(5). In fact, many other choice could also give extra 5 dimensional internal spaces while preserve different symmetry. As each choice will count unequal times for $x_i$ it seems that each choice will give different physical.  So, let us turn to the following observation.
\\
\\
{\bf Furthermore observation}.  Beside (2.2) and (2.3) we define
$$ [A,B]_{(zw)}= \partial_z A(x,y,z,w) \partial_w B(x,y,z,w)-\partial_w A(x,y,z,w) \partial_z B(x,y,z,w),\eqno{(2.14)}$$
$$ [A,B]_{(wx)}= \partial_w A(x,y,z,w) \partial_x B(x,y,z,w)-\partial_x A(x,y,z,w) \partial_w B(x,y,z,w),\eqno{(2.15)}$$
which also automatically satisfies the Jacobi identity.   Then, it can furthermore be proved that the representation
$$ [A(x,y,z,w),B(x,y,z,w)]= \sum_{\mu=x,y,z,w \atop\nu=x,y,z,w}\epsilon^{\mu\nu}\partial_\mu A(x,y,z,w) \partial_\nu B(x,y,z,w)$$
$$= [A,B]_{(xy)}+[A,B]_{(yz)}+[A,B]_{(zw)}+[A,B]_{(wx)},\eqno{(2.16)}$$
also satisfies the Jacobi identity.  In this case of 2-algebra we see that it could  appearing derivatives more than 3 coordinates.  Therefore, in case of  $(\mu,\nu,\lambda)=(x_1, x_2, x_3,x_4,x_5)$  and  
$$[f(x_1,x_2,x_3,x_4,x_5),g(x_1,x_2,x_3,x_4,x_5),h(x_1,x_2,x_3,,x_4,x_5)]=\epsilon^{\mu\nu\lambda}\partial_\mu f \partial_\nu g \partial_\lambda h,\eqno{(2.17)}$$
satisfies the fundamental identity then we will have a desired dimension of  internal space coming from 3-algebra to have KK6 from M2.  To see this let us prove the relevant theorem below.

\subsection{Theorem 2}
{\bf Theorem 2 }:  {\bf Define Nambu-Poisson n-bracket $ [f_1,\cdot\cdot\cdot,f_{n}]\equiv \epsilon^{i_1\cdot\cdot\cdot i_{n}} (\partial_{i_1} f_1\cdot\cdot\cdot\partial_{i_n} f_{n})$ then the fundamental identity could  be satisfied up to a total derivative in arbitrary space dimension}, i.e. 
$$[f_1,\cdot\cdot\cdot,f_{n-1},[g_1,\cdot\cdot\cdot,g_{n}]] = [g_2,\cdot\cdot\cdot,g_n,[f_1,\cdot\cdot\cdot, f_{n-1},g_1]]+\cdot\cdot\cdot+[g_1,\cdot\cdot\cdot,g_{n-1},[f_1,\cdot\cdot\cdot, f_{n-1},g_n]]$$
$$ - \epsilon^{i_1\cdot\cdot\cdot i_{n-1}[k} \epsilon^{j_1\cdot\cdot\cdot j_{n}]}\left( \partial_{j_1} g_1\cdot\cdot \cdot\partial_{j_{n}} g_{n}\right)\partial_k (\partial_{i_1} f_1\cdot\cdot\cdot\partial_{i_{n-1}} f_{n-1})$$
$$= [g_2,\cdot\cdot\cdot,g_n,[f_1,\cdot\cdot\cdot, f_{n-1},g_1]]+\cdot\cdot\cdot+[g_1,\cdot\cdot\cdot,g_{n-1},[f_1,\cdot\cdot\cdot, f_{n-1},g_n]]$$
$$-\partial_{k}\left(\epsilon^{i_1\cdot\cdot\cdot i_{n-1}[k} \epsilon^{j_1\cdot\cdot\cdot j_{n}]}\left( \partial_{j_1} g_1\cdot\cdot \cdot\partial_{j_{n}} g_{n}\right)(\partial_{i_1} f_1\cdot\cdot\cdot\partial_{i_{n-1}} f_{n-1})\right).\eqno{(2.18)}$$
\\
 {\it Proof} : Consider the arbitrary function $f_i$ and $g_j$ with antisymmetrization of $n+1$ indices $k$, $j_1,\cdot\cdot\cdot j_{n}$ 
$$ \epsilon^{i_1\cdot\cdot\cdot i_{n-1}[k} \epsilon^{j_1\cdot\cdot\cdot j_{n}]} 
 (\partial_{i_1} f_1\cdot\cdot\cdot\partial_{i_{n-1}} f_{n-1})\partial_k \left(
 \partial_{j_1} g_1\cdot\cdot \cdot\partial_{j_{n}} g_{n}
\right)= 0.~~~~~~~~~(property~I)\eqno{(2.19)}$$
The zero value in the above equation could be easily seen as the partial derivative index $k$ is antisymmetric with partial derivative indices $j_1,\cdot\cdot\cdot j_{n}$. Above equation implies 
$$ [f_1,\cdot\cdot\cdot,f_{n-1},[g_1,\cdot\cdot\cdot,g_{n}]]\equiv \epsilon^{i_1\cdot\cdot\cdot i_{n-1}k} \epsilon^{j_1\cdot\cdot\cdot j_{n}}
 (\partial_{i_1} f_1\cdot\cdot\cdot\partial_{i_{n-1}} f_{n-1})\partial_k \left(
 \partial_{j_1} g_1\cdot\cdot\cdot \partial_{j_{n}} g_{n}
\right)$$
$$ = \left(\epsilon^{i_1\cdot\cdot\cdot i_{n-1}j_1} \epsilon^{k j_2\cdot\cdot\cdot j_{n}} + \epsilon^{i_1\cdot\cdot\cdot i_{n-1}j_2} \epsilon^{j_1k\cdot\cdot\cdot j_{n}}  +\cdot\cdot\cdot+ \epsilon^{i_1\cdot\cdot\cdot i_{n-1} j_{n}} \epsilon^{j_1\cdot\cdot\cdot k}\right) (\partial_{i_1} f_1\cdot\cdot\cdot\partial_{i_{n-1}} f_{n-1})\partial_k \left( \partial_{j_1} g_1\cdot\cdot \cdot\partial_{n} g_{j_{n}}\right).\eqno{(2.20)}$$
Using the property that 
$$\epsilon^{k j_2\cdot\cdot\cdot j_{n}}\partial_k  \left(
 \partial_{j_2} g_2\cdot\cdot \cdot\partial_{j_{n}} g_{n}\right)=0.~~~~~~~~~(property~II),\eqno{(2.21)}$$
the first term in right-hand side of (2.20) becomes
$$\epsilon^{i_1\cdot\cdot\cdot i_{n-1}j_1} \epsilon^{k j_2\cdot\cdot\cdot j_{n}}(\partial_{i_1} f_1\cdot\cdot\cdot\partial_{i_{n-1}} f_{n-1})\partial_k \left( \partial_{j_1} g_1\cdot\cdot \cdot\partial_{j_{n}} g_{n}
\right)\hspace{3cm}$$
$$=\epsilon^{i_1\cdot\cdot\cdot i_{n-1}j_1} \epsilon^{k j_2\cdot\cdot\cdot j_{n}} (\partial_{i_1} f_1 \cdot\cdot\cdot \partial_{i_{n-1}} f_{n-1})  (\partial_{j_2} g_2 \cdot\cdot\cdot\partial_{j_n} g_n)(\partial_k \partial_{j_1} g_1)$$
$$ = \epsilon^{i_1\cdot\cdot\cdot i_{n-1}j_1} \epsilon^{k j_2\cdot\cdot\cdot j_{n}} (\partial_{j_2} g_2 \cdot\cdot\cdot\partial_{j_n} g_n) \partial_{k}\left(\partial_{i_1} f_1 \cdot\cdot\cdot \partial_{i_{n-1}} f_{n-1}   \partial_{j_1}g_1 \right) $$
$$- \epsilon^{i_1\cdot\cdot\cdot i_{n-1}j_1} \epsilon^{k j_2\cdot\cdot\cdot j_{n}}(\partial_{j_1}g_1 \cdot\cdot\cdot\partial_{j_n} g_n)\partial_k\left(\partial_{i_1} f_1 \cdot\cdot\cdot \partial_{i_{n-1}} f_{n-1} \right)$$
$$=[g_2,\cdot\cdot\cdot g_n,[f_1,\cdot\cdot\cdot f_{n-1},g_1]]-\epsilon^{i_1\cdot\cdot\cdot i_{n-1}j_1} \epsilon^{k j_2\cdot\cdot\cdot j_{n}}(\partial_{j_1}g_1 \cdot\cdot\cdot\partial_{j_n} g_n)\partial_k\left(\partial_{i_1} f_1 \cdot\cdot\cdot \partial_{i_{n-1}} f_{n-1} \right).\eqno{(2.22)}$$
In a same way we can find the similar relations for the second and third terms in right-hand side of (2.20). Collect the results we finally find the relation
$$ [f_1,\cdot\cdot\cdot, f_{n-1},[g_1,\cdot\cdot\cdot, g_{n}]]$$
$$= [g_2,\cdot\cdot\cdot,g_n,[f_1,\cdot\cdot\cdot, f_{n-1},g_1]]+\cdot\cdot\cdot+[g_1,\cdot\cdot\cdot,g_{n-1},[f_1,\cdot\cdot\cdot, f_{n-1},g_n]] $$
$$ - \left(\epsilon^{i_1\cdot\cdot\cdot i_{n-1}j_1} \epsilon^{k j_2\cdot\cdot\cdot j_{n}} + \epsilon^{i_1\cdot\cdot\cdot i_{n-1}j_2} \epsilon^{j_1k\cdot\cdot\cdot j_{n}}  +\cdot\cdot\cdot+ \epsilon^{i_1\cdot\cdot\cdot i_{n-1} j_{n}} \epsilon^{j_1\cdot\cdot\cdot k}\right)\left( \partial_{j_1} g_1\cdot\cdot \cdot\partial_{j_{n}} g_{n}\right)\partial_k (\partial_{i_1} f_1\cdot\cdot\cdot\partial_{i_{n-1}} f_{n-1}) .\eqno{(2.23)}$$
It is important to see that the  minus term can be expressed as 
$$\left( \epsilon^{i_1\cdot\cdot\cdot i_{n-1}[k} \epsilon^{j_1\cdot\cdot\cdot j_{n}]}- \epsilon^{i_1\cdot\cdot\cdot i_{n-1}k} \epsilon^{j_1\cdot\cdot\cdot j_{n}}\right) \left( \partial_{j_1} g_1\cdot\cdot \cdot\partial_{j_{n}} g_{n}\right)\partial_k (\partial_{i_1} f_1\cdot\cdot\cdot\partial_{i_{n-1}} f_{n-1})$$
$$=\epsilon^{i_1\cdot\cdot\cdot i_{n-1}[k} \epsilon^{j_1\cdot\cdot\cdot j_{n}]}\left( \partial_{j_1} g_1\cdot\cdot \cdot\partial_{j_{n}} g_{n}\right)\partial_k (\partial_{i_1} f_1\cdot\cdot\cdot\partial_{i_{n-1}} f_{n-1}),\eqno{(2.24)}$$
with the help of property II in (2.21).  Using again the property I in  (2.19) we see that (2.24) becomes
$$\epsilon^{i_1\cdot\cdot\cdot i_{n-1}[k} \epsilon^{j_1\cdot\cdot\cdot j_{n}]}\left( \partial_{j_1} g_1\cdot\cdot \cdot\partial_{j_{n}} g_{n}\right)\partial_k (\partial_{i_1} f_1\cdot\cdot\cdot\partial_{i_{n-1}} f_{n-1})$$
$$= \epsilon^{i_1\cdot\cdot\cdot i_{n-1}[k} \epsilon^{j_1\cdot\cdot\cdot j_{n}]}\left(\left( \partial_{j_1} g_1\cdot\cdot \cdot\partial_{j_{n}} g_{n}\right)\partial_k (\partial_{i_1} f_1\cdot\cdot\cdot\partial_{i_{n-1}} f_{n-1})+(\partial_{i_1} f_1\cdot\cdot\cdot\partial_{i_{n-1}} f_{n-1})\partial_k \left( \partial_{j_1} g_1\cdot\cdot \cdot\partial_{j_{n}} g_{n}\right)  \right)$$
$$ =\partial_{k}\left(\epsilon^{i_1\cdot\cdot\cdot i_{n-1}[k} \epsilon^{j_1\cdot\cdot\cdot j_{n}]}\left( \partial_{j_1} g_1\cdot\cdot \cdot\partial_{j_{n}} g_{n}\right)(\partial_{i_1} f_1\cdot\cdot\cdot\partial_{i_{n-1}} f_{n-1})\right).\eqno{(2.25)}$$
Therefore, in the arbitrary  spaces the fundamental identity could  be satisfied up to a total derivative.  Thus we have the following results: 
\\
\\
{\it Application 4:} In the case of $n$-algebra with $n$ spaces then any antisymmetrization of more than $n$ indices gives zero, a trick that leads to the so-called ``Schouten identities" [7]. Thus  
$$\epsilon^{i_1\cdot\cdot\cdot i_{n-1}[k} \epsilon^{j_1\cdot\cdot\cdot j_{n}]}\left( \partial_{j_1} g_1\cdot\cdot \cdot\partial_{j_{n}} g_{n}\right)\partial_k (\partial_{i_1} f_1\cdot\cdot\cdot\partial_{i_{n-1}} f_{n-1}) =0, ~~~~ n-algebra~with~n~spaces. \eqno{(2.26)}$$
which is an exact relation and fundamental identity is satisfied [7].  This prove the decomposability [6] of the Nambu-Poisson bracket.
\\
\\
{\it Application 5:} In the case of  $n$-algebra with $n+1$ spaces then use above result and theorem 1 we see that the fundamental identity is also satisfied.  The case of {\it application 2} is that of n=3.
\\
\\
The total derivative in the case of  applications 4 and 5  is zero.  We have also checked  the case of $3$-algebra with $5$ spaces and find that the total derivative term is not zero.  However,  the fundamental identity shall be satisfied to have a supersymmetry property in BLG theory [2].  

\subsection{Supersymmetry in NB BLG theory}
 To solve the problem we see that the variation of potential term in (1.1) is $\delta (Tr[X^{I},X^{J},X^{K}]$ $[X^I,X^J,X^K]) \sim  Tr((\delta [X^{I},X^{J},X^{K}]) [X^I,X^J,X^K])$.  Now, let us recall the property that the Lie-n algebra structure constant is defined by
$$  [g^{a_1}, g^{a_2}, \cdot\cdot\cdot,g^{a_n}]= \epsilon^{\mu_1\mu_2\cdot\cdot\cdot\mu_n}\partial_{\mu_1} g^{a_1}\partial_{\mu_2} g^{a_2}\cdot\cdot\cdot\partial_{\mu_n}g^{a_n} = f^{a_1a_2\cdot\cdot\cdot a_n}_{~~~~~~~~~~d}~ g^d. \eqno{(2.27)}$$
Thus, corresponding to  (2.24) there is an ``extra" term from the variation of potential term, which we investigate in below. 
$$\epsilon^{i_1\cdot\cdot\cdot i_{n-1}[k} \epsilon^{j_1 j_2\cdot\cdot\cdot j_{n}]}(\partial_{j_1}g_{a_1} \cdot\cdot\cdot\partial_{j_n} g_{a_n})\partial_k\left(\partial_{i_1} g_{b_1} \cdot\cdot\cdot \partial_{i_{n-1}} g_{b_{n-1}} \right) \cdot [g^{a_1}, g^{a_2}, \cdot\cdot\cdot,g^{a_n}]$$ 
$$ =\epsilon^{i_1\cdot\cdot\cdot i_{n-1}[k} \epsilon^{j_1 j_2\cdot\cdot\cdot j_{n}]}(\partial_{j_1}g_{a_1} \cdot\cdot\cdot\partial_{j_n} g_{a_n})\partial_k\left(\partial_{i_1} g_{b_1} \cdot\cdot\cdot \partial_{i_{n-1}} g_{b_{n-1}} \right) \cdot  f^{a_1a_2\cdot\cdot\cdot a_n}_{~~~~~~~~~~d}~ g^d$$ 
$$= \partial_k\left(\epsilon^{i_1\cdot\cdot\cdot i_{n-1}[k} \epsilon^{j_1 j_2\cdot\cdot\cdot j_{n}]}(\partial_{j_1}g_{a_1} \cdot\cdot\cdot\partial_{j_n} g_{a_n})\left(\partial_{i_1} g_{b_1} \cdot\cdot\cdot \partial_{i_{n-1}} g_{b_{n-1}} \right) \cdot  f^{a_1a_2\cdot\cdot\cdot a_n}_{~~~~~~~~~~d}~ g^d\right) $$
$$ - \epsilon^{i_1\cdot\cdot\cdot i_{n-1}[k} \epsilon^{j_1 j_2\cdot\cdot\cdot j_{n}]}(\partial_{j_1}g_{a_1} \cdot\cdot\cdot\partial_{j_n} g_{a_n})\left(\partial_{i_1} g_{b_1} \cdot\cdot\cdot \partial_{i_{n-1}} g_{b_{n-1}} \right) \cdot  f^{a_1a_2\cdot\cdot\cdot a_n}_{~~~~~~~~~~d}~ \partial_kg^d, \eqno{(2.28)}$$
in which we have used the property II in (2.21).  Now, using (2.27) we see that the second term in (2.28) becomes
$$Tr\left(\epsilon^{i_1\cdot\cdot\cdot i_{n-1}[k} \epsilon^{j_1 j_2\cdot\cdot\cdot j_{n}]}(\partial_{j_1}g_{a_1} \cdot\cdot\cdot\partial_{j_n} g_{a_n})\left(\partial_{i_1} g^{b_1} \cdot\cdot\cdot \partial_{i_{n-1}} g^{b_{n-1}} \right) \cdot f^{a_1a_2\cdot\cdot\cdot a_n}_{~~~~~~~~~~d}~ \partial_k g^d\right)$$
$$= Tr\left(f^{b_1b_2\cdot\cdot\cdot b_{n-1}[d}_{~~~~~~~~~~~~~~a}~f^{a_1a_2\cdot\cdot\cdot a_n]}_{~~~~~~~~~~~b} f_{a_1a_2\cdot\cdot\cdot a_nd}~g^a~g^b\right)$$
$$=f^{b_1b_2\cdot\cdot\cdot b_{n-1}[d}_{~~~~~~~~~~~~~~a}~f^{a_1a_2\cdot\cdot\cdot a_n]}_{~~~~~~~~~~~b} f_{a_1a_2\cdot\cdot\cdot a_nd}~h_{ab}= constant, \eqno{(2.29)}$$
in which we have used the metric form $h^{ab}$ defined in (1.3).  As the field $\chi(y)$ in (1.6) is the field $g$ in (2.28) and  coordinate $y$ is  the extra  dimensional internal spaces coming from n-algebra [10], we thus find that the term  (2.29) only contributes a constant to the action of extra dimensional internal space part.   

In conclusion, even if the Nambu-Poisson n-bracket could satisfy the fundamental identity only up to a total derivative in arbitrary space dimension the BLG Lagrangian still has invariant property up to a finite total derivative term, and NB BLG action could preserve the supersymmetry.  This means that we could have extra 5 dimensional internal spaces from 3-algebra in M2 brane to have 6 space dimensions in the world volume of KK6.  And, in general we could have arbitrary extra dimensional internal space from 3-algebra in M2 brane in BLG theory.
\section {U(1) Field on KK6 from BLG}
It is known that the M5 world-volume action contains self-dual 2-form gauge fields. In M5 form M2 [10] it is found that the self-dual 2-form gauge fields $A_{\mu\nu}$ in M5 could be produced from 1-form gauge fields $A_{\mu  b}^a$ in M2 BLG theory.  
\\

Now, as the KK6 world-volume action contains U(1) 1-form gauge fields [14] we have to show how the gauge field (denoted as $\tilde A_{\mu}$)  could be produced from $A_{\mu  b}^a$ field  in BLG theory.  For complete we will detail our approach in below.
\subsection{U(1) Field from Scalar Field Potential}
We show in this subsection how some fields on KK6 could be produced from  scalar field potential $Tr[X^{I},X^{J},X^{K}]^2$ in BLG.
\\

 $\bullet$  First, the scalar field $X^I$ and gauge field $A_{\mu ab}$ in BLG are expanded in terms of a basis $\chi^a(y)$ of  Nambu-Poisson bracket 
$$X^I(x,y) = \sum_{a}X_a^I(x)\chi^a(y).\eqno{(3.1)}$$
$$A_{\mu a}(x,y) = \sum_{a}A_{\mu ab}(x)\chi^b(y).\eqno{(3.2)}$$
as that in the case of M5 from M2 [10].  The coordinate $x$ is the 3 dimensional  world volume of M2 brane while coordinate $y$ is the extra 4 dimensional internal spaces coming from NB Lie 3-algebra.  

The theorem shown in section II tells us that we can have sufficiently extra dimension to allow  KK6 from M2.  Thus, totally we could have 7 dimensional spacetimes of KK6.  We use $\mu$, $\nu$, $\lambda$ to label the longitudinal coordinate $x$ of the M2 worldvolume, which contain the isometry direction $z$.  We also use $I$, $J$, $K$ to label the transverse directions to the M2  worldvolume. 
\\

$\bullet$ Next, the index $I(=1,...,8)$ is decomposed as following.  We use  $\dot\mu$, $\dot\nu$, $\dot\lambda$ (=4,...,8) to label the longitudinal directions (coordinate $y$) to the KK6 worldvolume, and $i$, $j$, $k$ (=1,2,3) to label the  transverse directions to the KK6 worldvolume, which shall not contain the isometry direction $z$, as mentioned in section one.   Therefore the potential term in (1.1) is decomposed as 
$$ Tr[X^{I},X^{J},X^{K}][X^I,X^J,X^K] =Tr([X^{\dot\mu},X^{\dot\nu},X^{\dot\lambda}])^2+3~ Tr([X^{\dot\mu},X^{\dot\nu},X^{i}])^2$$
$$+3~Tr([X^{\dot\mu},X^{i},X^{j}])^2+Tr([X^{i},X^{j},X^{k}])^2  \eqno{(3.3)}$$

$\bullet$  Now, we consider a  fluctuation field $\tilde A^{\dot\mu}$ appearing on the  3-algebra coordinate $y$ and expand the field $X^{\dot\mu}(x,y)$ by
$$X^{\dot\mu}(x,y)=  y^{\dot\mu}+ \tilde A^{\dot\mu}(x,y),\eqno{(3.4)}$$
as that in [10].  The NB Lie-3 algebra is defined by
$$  [\chi^{a}, \chi^{b},\chi^{c}]= \sum_{\dot\mu\dot\nu\dot\lambda}\epsilon^{\dot\mu\dot\nu\dot\lambda}\partial_{\dot\mu} \chi^{a}\partial_{\dot\nu} \chi^{b}\partial_{\dot\lambda} \chi^{c}= f^{abc}_{~~~d}~\chi^d, \eqno{(3.5)}$$
and we have a relation 
$$  [y^{\dot\mu}, y^{\dot\nu},y^{\dot\lambda}]= \epsilon^{\dot\mu\dot\nu\dot\lambda}. \eqno{(3.6)}$$

Using the properties of (3.5) and (3.6) we can find that 
$$Tr([X^{\dot\mu},X^{\dot\nu},X^{\dot\lambda}])^2= Tr\Big(\epsilon^{\dot a\dot b \dot c}\partial_{\dot a}X^{\dot\mu}\partial_{\dot b}X^{\dot\nu}\partial_{\dot c}X^{\dot\lambda}\Big)^2\hspace{6.5cm}$$
$$\hspace{2.5cm}= Tr\Big(\epsilon^{\dot a\dot b \dot c}~\partial_{\dot a}[ y^{\dot\mu}+ \tilde A^{\dot\mu}(x,y)]\partial_{\dot b}[ y^{\dot\nu}+ \tilde A^{\dot\nu}(x,y)]\partial_{\dot c}[ y^{\dot\lambda}+ \tilde A^{\dot\lambda}(x,y)]\Big)^2$$
$$=(\partial^{\dot\mu}\tilde A^{\dot\nu})^2+\cdot\cdot\cdot.\hspace{5.3cm}\eqno{(3.7)}$$
The dot term in (3.7) is a constant (coming from $(\epsilon^{\dot\mu\dot\nu\dot\lambda})^2$) or the terms linear in $\partial \tilde A^{\dot\mu}(x,y)$ which becomes zero after integration by part in the action.  The term $O(\tilde A^3)$ are also neglected as we consider only the  quadratic terms of  $\tilde A^{\dot\mu}(x,y)$  behavior in  this paper. 

  In the same way we find that
$$Tr([X^{\dot\mu},X^{\dot\nu},X^{i}])^2=2(\partial^{\dot\mu}X^i)^2+\cdot\cdot\cdot.\hspace{7.4cm}\eqno{(3.8)}$$
The 3rd and 4th terms in (3.3) have no quadratic terms of $\tilde A^{\dot\mu}$ and are also neglected.  
\\

$\bullet$ Note that in [10] it defines $ \epsilon_{\dot\mu\dot\nu\dot\lambda}\tilde A^{\dot\lambda}\equiv A_{\dot\mu\dot\nu}$ which is identified as a part of self-dual 2-form gauge fields on M5.  In this paper we let $\tilde A^{\dot\lambda}$ itself as a part of U(1) gauge fields on KK6. 

\subsection{U(1) Field from Chern-Simon Term}
We show in this subsection how some fields on KK6 could be produced from Chern-Simons term of  1-form gauge field $A_{\mu ab}$ in BLG.
\\

$\bullet$  First, using the definition (3.2) the Chern-Simons term in (1.1) can be rewritten as
$$ L_{CS}=\frac{1}{2}\epsilon^{\mu\nu\lambda}([[\chi^a,\chi^b,\chi^c] , \chi^d] A_{\mu a b}\partial_\nu A_{\lambda c d}+\frac{2}{3} [[\chi^c, \chi^d, \chi^a
], \chi_g][[\chi^e, \chi^f,\chi^g], \chi^b]]A_{\mu a b}A_{\nu c d} A_{\lambda ef} \Big)$$
$$= \frac{1}{2}\epsilon^{\mu\nu\lambda}\Big(
[[A_{\mu b} ,\chi^b, \partial_\nu A_{\lambda d}], \chi^d]+\frac{2}{3}[[
A_{\nu d}, \chi^d, A_{\mu b}],\chi_g][[A_{\lambda f},\chi^f, \chi^g],\chi^b]]\Big).\eqno{(3.9)}$$
~

$\bullet$ Next, we pick the first five basis in (3.2) as coordinates, i.e.
$$\chi^{\dot\mu}=y^{\dot\mu}.\eqno{(3.10)}$$
The rest of the basis correspond to higher oscillations modes and are ignored, as that in [10].  

 Using the NB property of (3.5) the Chern-Simons term in (3.9) can be easily calculated.  The results are 
$$L_{CS}=\epsilon^{\mu\nu\lambda}\epsilon^{\dot\mu\dot\nu\dot\lambda}\partial_\mu A_{\nu\dot\mu}\partial_{\dot\nu} A_{\lambda\dot\lambda} + O(A^3) ,\eqno{(3.11)}$$
in which $A_{\mu\dot\nu}=A_{\mu ba}\chi^b$ while ${\dot\nu}={a}$ as we pick only first five basis.  To read the U(1) gauge field $\tilde A_{\mu}$ from (3.11) we need to use the isometric property of KK6 and is discussed in section 3.4. 

\subsection{U(1) Field from Scalar Field Kinetic Term}
We show in this subsection how some fields on KK6 could be produced from  scalar field kinetic term $(D^\mu X^{I})^2$ in BLG.
\\

$\bullet$  Using   (3.4) $\sim$ (3.6) and (3.10) the  kinetic term in (1.1) could be calculated as following [10].  
Using the relation
$$D_\mu X^{I}\equiv \partial_\mu (X^I_a \chi^a)-f^{cdb}_{~~~~a}A_{\mu cd}X^I_b \chi^a=\partial_\mu X^I-[\chi^c,\chi^d,\chi^b,]A_{\mu cd}X^I_b= \partial_\mu X^I-[A_{\mu d},\chi^d,X^I],\eqno{(3.12)}$$
we can find that
$$ D_\mu X^{\dot\nu}=\partial_\mu X^{\dot\nu}-[A_{\mu {\dot\lambda}},\chi^{\dot\lambda},X^{\dot\nu}] =\partial_\mu (y^{\dot\nu}+A^{\dot\nu})-[A_{\mu {\dot\lambda}},y^{\dot\lambda},y^{\dot\nu}]+\cdot\cdot\cdot \approx\partial_\mu A^{\dot\nu}- \epsilon^{\dot\mu\dot\lambda\dot\nu}\partial_{\dot\mu}A_{\mu\dot\lambda},\eqno{(3.13)}$$
in which we keep only the linear in $A$. 

  In the same way we find that 
$$ D_\mu X^{i}=\partial_\mu X^{i}-[A_{\mu {\dot\lambda}},\chi^{\dot\lambda},X^{i}] \approx \partial_\mu X^{i}.\eqno{(3.14)}$$
Thus
$$(D^\mu X^{I})^2=(\partial^{\mu}X^i)^2+(\partial_\mu \tilde A^{\dot\nu}-\epsilon^{\dot\mu\dot\nu\dot\lambda}\partial_{\dot\mu} A_{\mu\dot\lambda})^2.\eqno{(3.15)}$$
 Now, the total kinetic term of scalar field $X^i$ on the KK6 could be found in (3.8) and (3.15). 
\\

$\bullet$  Note that a part of  U(1) gauge fields kinetic term has been shown in (3.7).   What we lack is the terms $\partial_{\dot\mu}  \tilde A_{\nu}$, $\partial_{\mu}  \tilde A_{\dot\nu}$ and $\partial_{\mu}  \tilde A_{\nu}$ which shall come from the terms in (3.11) and (3.15) to obtain the U(1) Lagrangian on KK6.  We will  in next subsection see that  the isometry property of KK6 plays an important role to obtain the desired terms. 
\subsection{U(1) Field and Isometry of KK6}
~~~~~~~$\bullet$  First, we note that the indices $\mu$, $\nu$, $\lambda$ label the longitudinal coordinate of the M2 worldvolume, which contains the  isometry direction $z$.  However, as mentioned in section one, the KK monopole cannot move in this direction one should not associate a physical worldvolume scalar to it. Therefore we will first separate  the index $\mu$(=0,1,2) to $\tilde\mu$(=0,1)  and $z$.  Thus the worldvolume of KK6 has the five index $\dot\mu$(=3,..,8) plus 2 index $\tilde\mu$(=0,1).
\\

$\bullet$   Next, under the above decomposition of index the Chern-Simons term (3.11) becomes
$$\epsilon^{\mu\nu\lambda}\epsilon^{\dot\mu\dot\nu\dot\lambda}\partial_\mu A_{\nu\dot\mu}\partial_{\dot\nu} A_{\lambda\dot\lambda}= \epsilon^{\tilde\mu\tilde\nu}\epsilon^{\dot\mu\dot\nu\dot\lambda}\left[\partial_{z}A_{\tilde\mu\dot\mu}\partial_{\dot\nu} A_{\tilde\nu\dot\lambda}+\partial_{\nu}A_{z\dot\mu}\partial_{\dot\nu} A_{\tilde\mu\dot\lambda}+\partial_{\tilde\mu}A_{\tilde\nu\dot\mu}\partial_{\dot\nu} A_{z\dot\lambda}\right]$$
$$= - \epsilon^{\tilde\mu\tilde\nu}\partial_{\tilde\mu}A_{\tilde\nu\dot\mu}\partial^{\dot\mu} \tilde A_{z}\hspace{2.3cm}.\eqno{(3.16)}$$
To obtain above result we have performed integration by part and define 
$$\epsilon^{\dot\mu\dot\nu\dot\lambda}\partial_{\dot\mu}A_{z\dot\lambda}=\partial^{\dot\nu}\tilde A_{z}.\eqno{(3.17)}$$
We also let 
$$\partial_{z}A_{\tilde\mu\dot\mu}=0.\eqno{(3.18)}$$ 
This is because that  M2-brane can only intersect with  KK6  over a 0-brane such that one of the worldvolume directions of the M-2-brane coincides with the isometry direction z, which is a wrapped M2 brane [13], and thus the coordinate $z$ dose not belong to worldvolume of KK6.

Note that above procedure is actually performing a (worldvolume) dimensional reduction of the BLG theory in the z  direction, as made explicit by the Eq. (3.18) and the fact that one starts with 3 original + 5 emergent worldvolume directions and ends up with a 7-dimensional worldvolume, as expected for a KK6 in M-theory.
\\

$\bullet$  After decomposition the  kinetic term  (3.15) becomes 
$$(D^\mu X^{I})^2=(\partial^{\mu}X^i)^2+(\partial_{\tilde\mu} \tilde A^{\dot\nu}-\epsilon^{\dot\mu\dot\nu\dot\lambda}\partial_{\dot\mu} A_{{\tilde\mu}\dot\lambda})^2+(\partial_{z} \tilde A^{\dot\nu}-\epsilon^{\dot\mu\dot\nu\dot\lambda}\partial_{\dot\mu} A_{z\dot\lambda})^2$$
$$=(\partial^{\mu}X^i)^2+(\partial_{\tilde\mu} \tilde A^{\dot\nu}-\partial^{\dot\nu}\tilde A_{\tilde\mu})^2+(\partial^{\dot\nu}\tilde A_{z})^2,\hspace{1.3cm}\eqno{(3.19)}$$
in which we have used the relations (3.17) and (3.18).  We also define the field $\tilde A_{\tilde\mu}$ by the relation 
$$\epsilon^{\dot\mu\dot\nu\dot\lambda}\partial_{\dot\mu} A_{\tilde\mu\dot\lambda}\equiv \partial^{\dot\nu}\tilde A_{\tilde\mu}.\eqno{(3.20)}$$
The field $\tilde A_z$ in (3.19) is special and need a careful treatment.
\\ 

$\bullet$  To proceed, we know that as the $z$ is the isometry direction the coordinate $z$ dose not belong to worldvolume of KK6.  Thus the field $\tilde A_z$ in here is not a dynamic field on KK6 worldvolume and shall be eliminated.  This can be done as following.

    First, we substitute (3.16) into the Chern-Simons term in BLG Lagrangian (1.1) and substitute (3.19) into the  kinetic term in the BLG Lagrangian (1.1). The  Lagrangian then has the following terms

$$L = -{1\over2}(\partial^{\dot\nu}\tilde A_{z})^2- \epsilon^{\tilde\mu\tilde\nu}\partial_{\tilde\mu}A_{\tilde\nu\dot\mu}\partial^{\dot\mu} \tilde A_{z}+\cdot\cdot\cdot.\eqno{(3.21)}$$

 Next, take a variation with respective to $\partial^{\dot\nu}\tilde A_{z}$ we can find the solution of field $\partial^{\dot\nu}\tilde A_{z}$.   The result is 
$$\partial^{\dot\nu}\tilde A_{z}= -\epsilon^{\tilde\mu\tilde\nu}\partial_{\tilde\mu}A_{\tilde\nu\dot\mu}.\eqno{(3.22)}$$

Finally, substitute the above solution into  (3.21) we therefore can eliminate $\tilde A_z$ and find that the Lagrangian contains the term 
$${L} \sim (\epsilon^{\tilde\mu\tilde\nu}\partial_{\tilde\mu}A_{\tilde\nu\dot\mu}) (\epsilon^{\tilde\lambda\tilde\delta}\partial_{\tilde\lambda}A_{\tilde\delta}^{~\dot\mu})+\cdot\cdot\cdot.\eqno{(3.23)}$$
~

$\bullet $ Now we come to the final step.  We define above term as 
$$(\epsilon^{\tilde\mu\tilde\nu}\partial_{\tilde\mu}A_{\tilde\nu\dot\mu}) (\epsilon^{\tilde\lambda\tilde\delta}\partial_{\tilde\lambda}A_{\tilde\delta}^{~\dot\mu})\equiv (\partial_{\tilde\mu}\tilde A_{\tilde\nu}-\partial_{\tilde\nu}\tilde A_{\tilde\mu})(\partial^{\tilde\mu}\tilde A^{\tilde\nu}-\partial^{\tilde\nu}\tilde A^{\tilde\mu}).\eqno{(3.24)}$$
Solve (3.23) we can know how $\tilde A_{\tilde\mu}$ depends on coordinate ${\tilde\nu}$.  And solve (3.20) we can know how $\tilde A_{\tilde\mu}$ depends on coordinate ${\dot\nu}$.   Therefore, we can find the U(1) gauge fields $\tilde A_{\tilde\mu}(x,y)$ from the gauge field $A_{\mu  b}^a$ field in BLG theory and obtain the desired U(1) Lagrangian from the BLG action.  These complete our investigations.
\section {Discussion}
In this paper, we consider that the BLG  is an universal M2 theory. Then BLG action shall be able to produce worldvolume action of  more extended objects besides M5 [15].  We first establish the general theorem to see that we could have sufficient extra dimensional internal spaces from 3-algebra in M2 brane to have 6 space dimensions in the world volume of KK6.  We next use the special property of  isometry direction in  $(0|M2,KK6)$ to find the U(1) field on KK6 world-volume action.   Thus, we have found a possibility that the Kaluza-Klein monopole (KK6) world-volume action may be obtained from the multiple M2 action which is described by  BLG theory.   

Finally, as there is the configuration  $(1|M2,M9)$ [11] it will be interesting to derive the M9 action from M2.  In this case as the M9 is a massive brane [16] it needs to carefully deal with.   Note that the action of the multiple M2-branes in M-theory was also found by Aharony, Bergman, Jafferis and Maldacena (ABJM) [17] after the ground-breaking works of BLG theory.  The problem of KK6 from M2 in ABJM theory is also deserved to be investigated [18]. 
\\
\\
\\
\\
{\bf Acknowledgments} :  The author thanks Kuo-Wei Huang for interesting and useful discussions about M theory and KK6 properties. This work is supported in part by the Taiwan National Science Council. 
\\
\\
\\
\begin{center} {\bf REFERENCES}\end{center}
\begin{enumerate}
\item J. Bagger and N. Lambert, ``Modeling multiple M2s", Phys. Rev. D 75 (2007) 045020  [arXiv:hep-th/0611108].
\item J. Bagger and N. Lambert,``Gauge Symmetry and Supersymmetry of Multiple M2-Branes", Phys. Rev. D 77 (2008) 065008  [arXiv:0711.0955 [hep-th]]. 
\item J. Bagger and N. Lambert, ``Comments On Multiple M2-branes", JHEP 0802 (2008) 105  [arXiv:0712.3738 [hep-th]].
\item A. Gustavsson, ``Algebraic structures on parallel M2-branes", Nucl.Phys.B811 (2009) 66  [arXiv:0709.1260 [hep-th]].
\item Y. Nambu, ``Generalized Hamiltonian dynamics", Phys. Rev. D 7 (1973) 2405 . 
\item L. Takhtajan, ``On Foundation Of The Generalized Nambu Mechanics (Second Version)", Commun. Math. Phys. 160 (1994) 295  [arXiv:hep-th/9301111];\\ D. Alekseevsky, P. Guha, ``On Decomposability of Nambu-Poisson Tensor", Acta. Math. Univ. Commenianae 65 (1996) 1.
\item V. T. Filippov, ``n-Lie algebras", Sib. Mat. Zh.,26, No. 6 (1985) 126140;\\  Jose A. de Azcarraga, Jose M. Izquierdo ``n-ary algebras: a review with applications" [arXiv:1005.1028 [math-ph]].
\item M. Van Raamsdonk, ``Comments on the Bagger-Lambert theory and multiple M2-branes", JHEP0805 (2008) 105 [arXiv:0803.3803 [hep-th]]; \\N. Lambert and D. Tong, ``Membranes on an Orbifold", Phys.Rev.Lett.101 2008 041602 [arXiv:0804.1114 [hep-th]];\\ J. Distler, S. Mukhi, C. Papageorgakis and M. Van Raamsdonk, ``M2-branes on M-folds", JHEP 0805 (2008) 038 [arXiv:0804.1256 [hep-th].
\item P.-M. Ho, R.-C. Hou, and Y. Matsuo, ``Lie 3-algebra and multiple M2-branes", JHEP 06 (2008) 020, [arXiv:0804.2110 [hep-th]].
\item P.-M. Ho and Y. Matsuo, ``M5 from M2", JHEP 06 (2008) 105, [arXiv:0804.3629 [hep-th]];\\ P.-M. Ho, Y. Imamura, Y. Matsuo, and S. Shiba, ``M5-brane in three-form flux and multiple M2-branes", JHEP 08 (2008) 014, [arXiv:0805.2898 [hep-th]].
\item C.M. Hull, ``Gravitational Duality, Branes and Charges", Nucl. Phys. B509 (1998) 216, [hep-th/9705162];\\  E. Bergshoeff, J. Gomis, P. K. Townsend, ``M-brane intersections from worldvolume superalgebras," Phys.Lett. B421 (1998) 109 [hep-th/9711043].
\item  R.D. Sorkin, Phys. Rev. Letters 51 (1983) 87;\\ D.J. Gross and M. Perry, Nucl. Phys. B226 (1983) 29.
\item  E. Bergshoeff, M. de Roo, E. Eyras, B. Janssen and J.P. van der Schaar, ``Intersections involving monopoles and waves in eleven dimensions", Class. Quantum Grav. 14 (1997) 2757, [hep-th/9704120];\\ E. Bergshoeff,  E. Eyras,  and Y. Lozano, ``The Massive Kaluza-Klein Monopole", Phys.Lett. B430 (1998) 77-86 [hep-th/9802199].
\item E. Bergshoeff, B. Janssen and T. Ortin, ``Kaluza-Klein Monopoles and Gauged Sigma-Models", Phys. Lett. B410 (1997) 132 [hep-th/9706117].
\item  C. Krishnan and C. Maccaferri, ``Membranes on Calibrations ," JHEP 0807 (2008) 005 [arXiv:0805.3125[hep-th]]. 
\item  E. Bergshoeff, Y. Lozano, T. Ortin, ``Massive Branes ," Nucl.Phys. B518 (1998) 363 [hep-th/9712115].
\item O. Aharony, O. Bergman, D. L. Jafferis and J. Maldacena, ``N=6 superconformal Chern-Simons matter theories, M2-branes and their gravity duals," JHEP 0810 (2008) 091  [arXiv: 0806.1218 [hep-th]].
\item  Wung-Hong Huang, in preparation.
\end{enumerate}
\end{document}